\documentclass[a4paper,11pt]{article}
\pdfoutput=1 % if your are submitting a pdflatex (i.e. if you have
\usepackage{jinstpub} % for details on the use of the package, please
\title{\boldmath Analysis of the light production and propagation in the 4-tonne dual-phase demonstrator}

\author{Chiara Lastoria}
\affiliation{CIEMAT, Centro de Investigaciones Energéticas, Medioambientales y Tecnológicas,\\Av. Complutense 40, 28040 Madrid, Spain}

\emailAdd{chiara.filomena.lastoria@cern.ch}

\abstract{\\The  Deep  Underground  Neutrino  Experiment  (DUNE)  is  a  leading-edge experiment designed to perform neutrino science and proton decay searches. In particular, the far detector  will  consist  of  four  10-kton Liquid Argon (LAr) Time Projection Chambers using  both  single  and  dual-phase technologies. The  latter  provides  charge  amplification in the gaseous phase. In order to optimize these designs, two large  prototypes are  taking data at CERN since 2018. Previously, a dual-phase 4-tonne demonstrator was constructed and exposed to cosmic muons in 2017 and exhibited good performance in terms of charge and light collection. 

The light detection system is important to provide a trigger to the charge acquisition system and to obtain additional information from the scintillation light produced in the particle interaction. In the demonstrator, five cryogenic photo-multipliers were installed with different base polarity configurations and wavelength shifting methods. During the detector operation, scintillation light data were collected in different drift and amplification field conditions. An overview of the light detection system performance and results on the light production and propagation are presented. Our studies allowed to improve the understanding of some LAr properties.}

\arxivnumber{1911.06880} % only if you have one

\collaboration[c]{on behalf of DUNE collaboration}

\proceeding{LIDINE 2019: LIght Detection In Noble Elements\\
28-30 August 2019\\
University of Manchester}

\begin{document}
\maketitle
\flushbottom

%For internal references use label-refs: see section~\ref{sec:intro}.
%Bibliographic citations can be done with cite: refs.~\cite{a,b,c}.
%A similar solution is available for figures via the \texttt{subfigure}
%package (not loaded by default and not shown here).
%See figure~\ref{fig:i} and table~\ref{tab:i}.

\section{Introduction}
\label{sec:intro}
The next generation neutrino experiment DUNE \cite{DUNE_IDR} primarily aims to investigate CP-violation in the leptonic sector and to study the neutrino mass hierarchy. To pursue these goals, DUNE plans to build a far detector made up of four Time Projection Chambers (TPCs) filled with 10-kton of Liquid Argon (LAr). The detector design consists of both single and dual-phase (DP) technologies. The DUNE physics program also includes proton decay searches and study of low energy neutrinos from supernova core-collapse.

In these detectors, a particle crossing the LAr volume produces ionization and excitation of the Ar atoms. The presence of an electric field suppresses possible recombination processes and allows the drift of the ionization electrons collected on a finely segmented anode plane. In the DP configuration, while longer drift paths are possible, higher signal-to-noise ratio can be achieved thanks to the signal amplification in the Gas Argon (GAr) phase; there the electrons are extracted and multiplied by a stronger electric field in the Large Electron Multiplier (LEM) holes. In addition, complementary information is obtained from two benchmark light signals, the prompt scintillation light (S1), produced in the LAr phase, and the electro-luminescence light (S2), from the electrons acceleration in GAr amplification region. This light, generated at $\sim$128 nm, is typically collected by photo-multipliers tubes (PMTs) where a wavelength shifter makes its detection possible. More details on the light photon detection systems in the DUNE DP design can be found here \cite{DUNEDP_IDR}.

\section{Detector description, light detection system and trigger systems}
\label{sec:det}
The scalability of the DP technology toward the kiloton-scale far detector modules is being demonstrated through the performance of two prototypes built at CERN: the 4-tonne demonstrator \cite{4TON} exposed to cosmic rays in 2017, with a LAr fiducial volume of 3$\times$1$\times$1 m$^3$, and the 300-tonne ProtoDUNE-DP detector \cite{ProtoDUNEDP}, with a fiducial volume of 6$\times$6$\times$6 m$^3$ of LAr, currently under commissioning. The operation of the 4-tonne demonstrator had the important role to prove the feasibility of such technology at the ton scale. It demonstrated the capability to drift the electrons over 1 m, and to extract, amplify and collect them on a 3 m$^2$ charge readout plane (CRP). The CRP consisted of 50x50 cm$^2$ LEM-anode units pre-assembled together. 

\begin{figure}[hptb]
  \centering 
  \includegraphics[width=0.19\linewidth, height=4.2cm ]{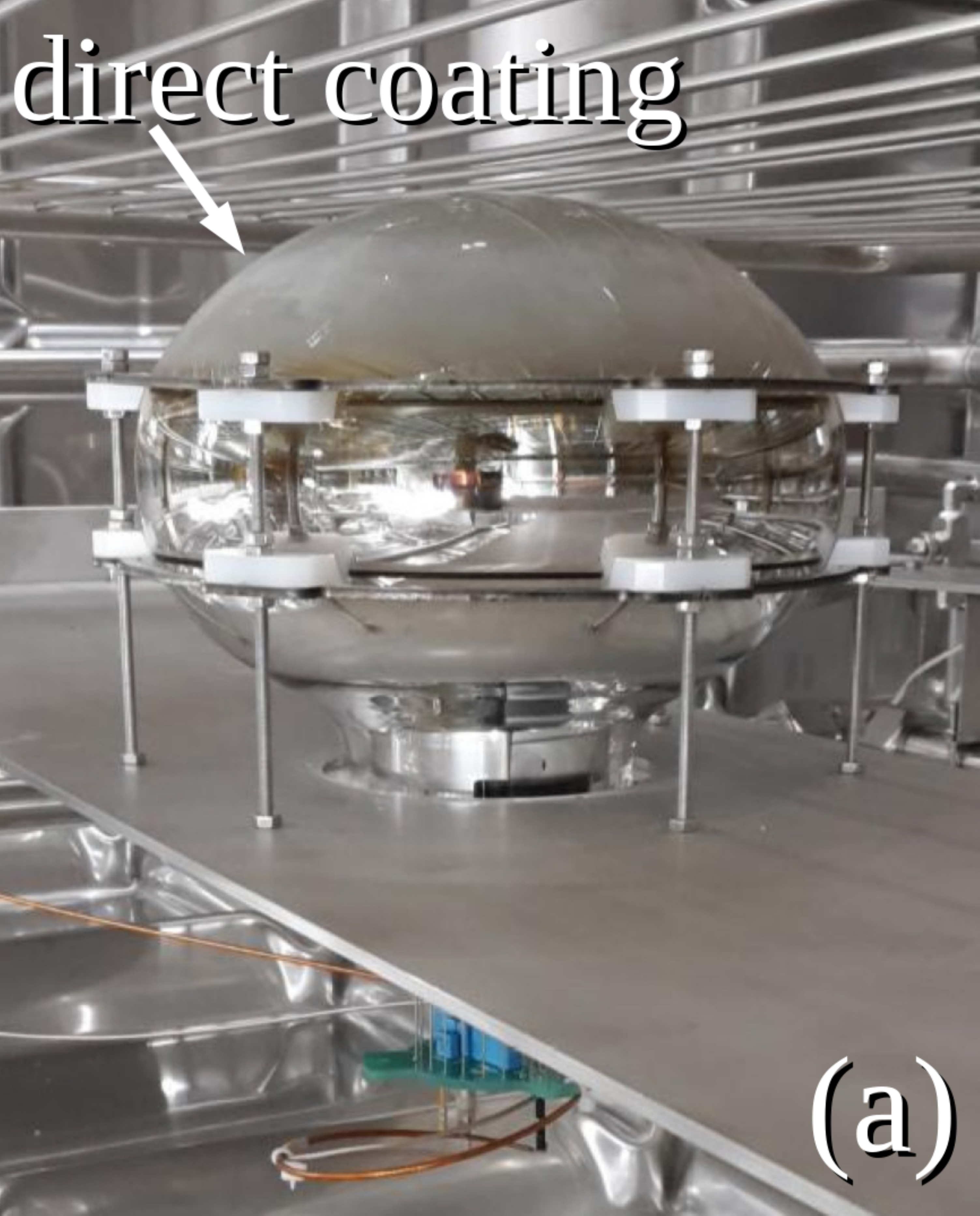}
  %\qquad
  \includegraphics[width=0.19\linewidth, height=4.2cm ]{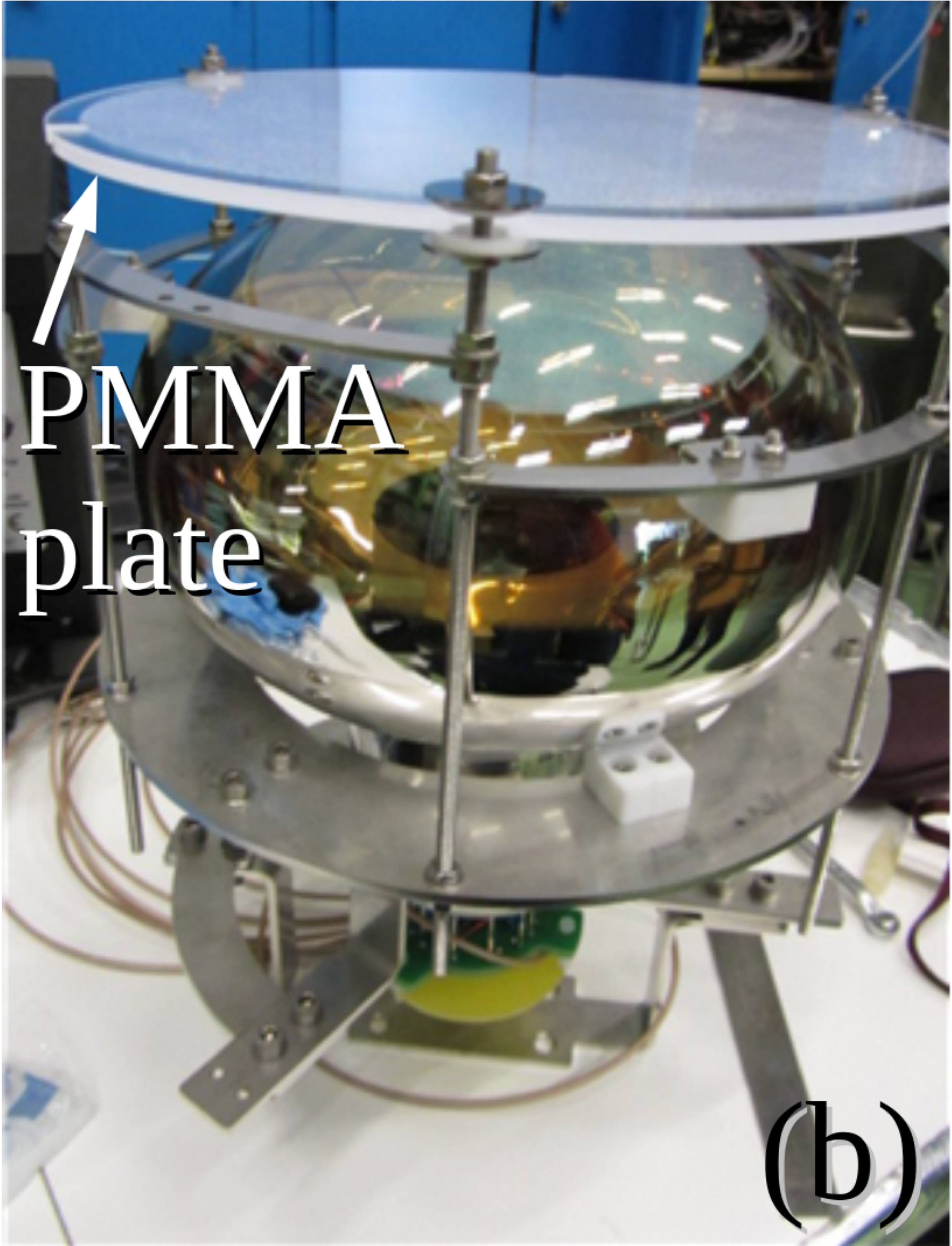}
  \qquad
  \includegraphics[width=0.35\linewidth, height=4.2cm ]{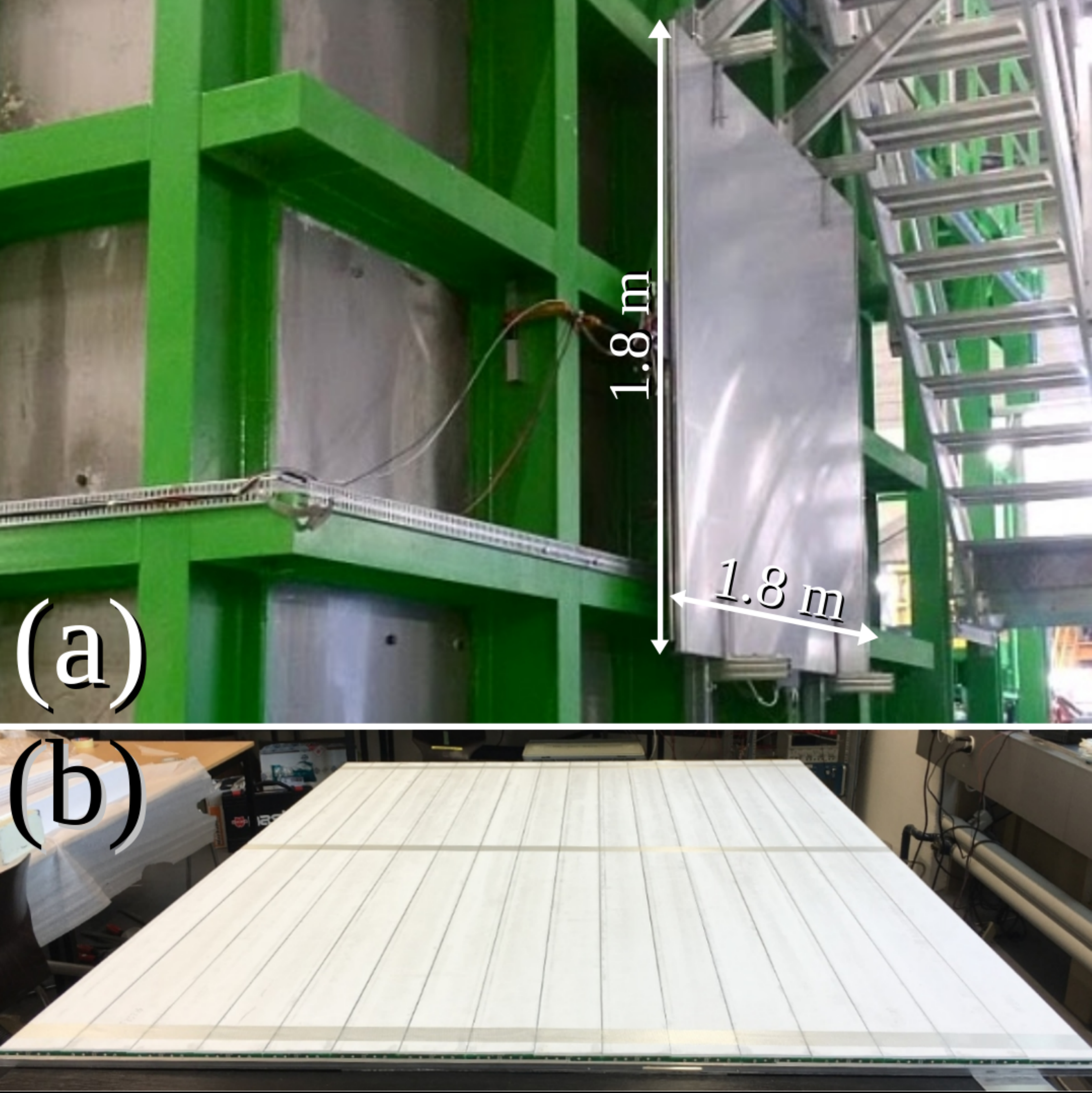}
  \caption{\textbf{Left}: Picture of the two TPB coating options used as a wavelength shifter in the 4-tonne demonstrator: direct coating over the PMT photo-cathode \textbf{(a)} or over a poly-methyl methacrylate (PMMA) plate \textbf{(b)}. \textbf{Right}: Picture of a pair of CRT panels positioned outside the cryostat walls \textbf{(a)}, each panel is made by 16 strips to allow tracks reconstruction \textbf{(b)}, the picture is taken from \cite{CRT_paper}.\label{fig:311pictures}  }
\end{figure}

The light detection system was composed of five eight inches cryogenic R5912-02Mod
\footnote{http://pdf1.alldatasheet.com/datasheet-pdf/view/397633/HAMAMATSU/R5912-02.html} 
PMTs \cite{PMTs_ref} positioned behind the cathode. The performance of different PMT-base polarity configurations (positive and negative bases) and wavelength shifter coatings were tested. Tetraphenyl-butadiene (TPB) was selected as wavelength shifter and applied using two different methods: a direct coating over the PMT photo-cathode or over a poly-methyl methacrylate (PMMA) plate installed above the PMT surface, both shown in figure \ref{fig:311pictures} - left. 

In order to trigger on crossing muons, two alternative systems were developed: a PMT self-trigger and an external Cosmic Rays Tagger (CRT) trigger. The self-trigger was based on a five-fold coincidence of the prompt scintillation light signal from the 5 PMTs, providing the $\mathrm{t_{0}}$ time for the charge DAQ. The CRT trigger was given by two pairs of 1.8x1.8 m$^{2}$ scintillator panels \cite{CRT_paper} positioned outside the cryostat, figure \ref{fig:311pictures} - right. Due to the geometrical acceptance of the CRT panels, the typical trigger rate was around $\sim$ 0.3 Hz. 

\section{Light signal characterization}
\label{sec:lightana}
A detailed analysis of the prompt scintillation light signal has been performed considering the light yield and the scintillation time profile features in different detector conditions. The characterization of the electro-luminescence light was possible thanks to the development of a dedicated algorithm to reconstruct the S2 signal. The study of the S1 and S2 signals allowed to retrieve additional key information on the light produced by the muons crossing the demonstrator and its propagation inside the LAr active volume. 

\subsection{Study of the Rayleigh scattering length}
\label{sec:Ray}

\begin{figure}[h!ptb]
 \centering 
 \includegraphics[width=0.45\linewidth, height=5cm]{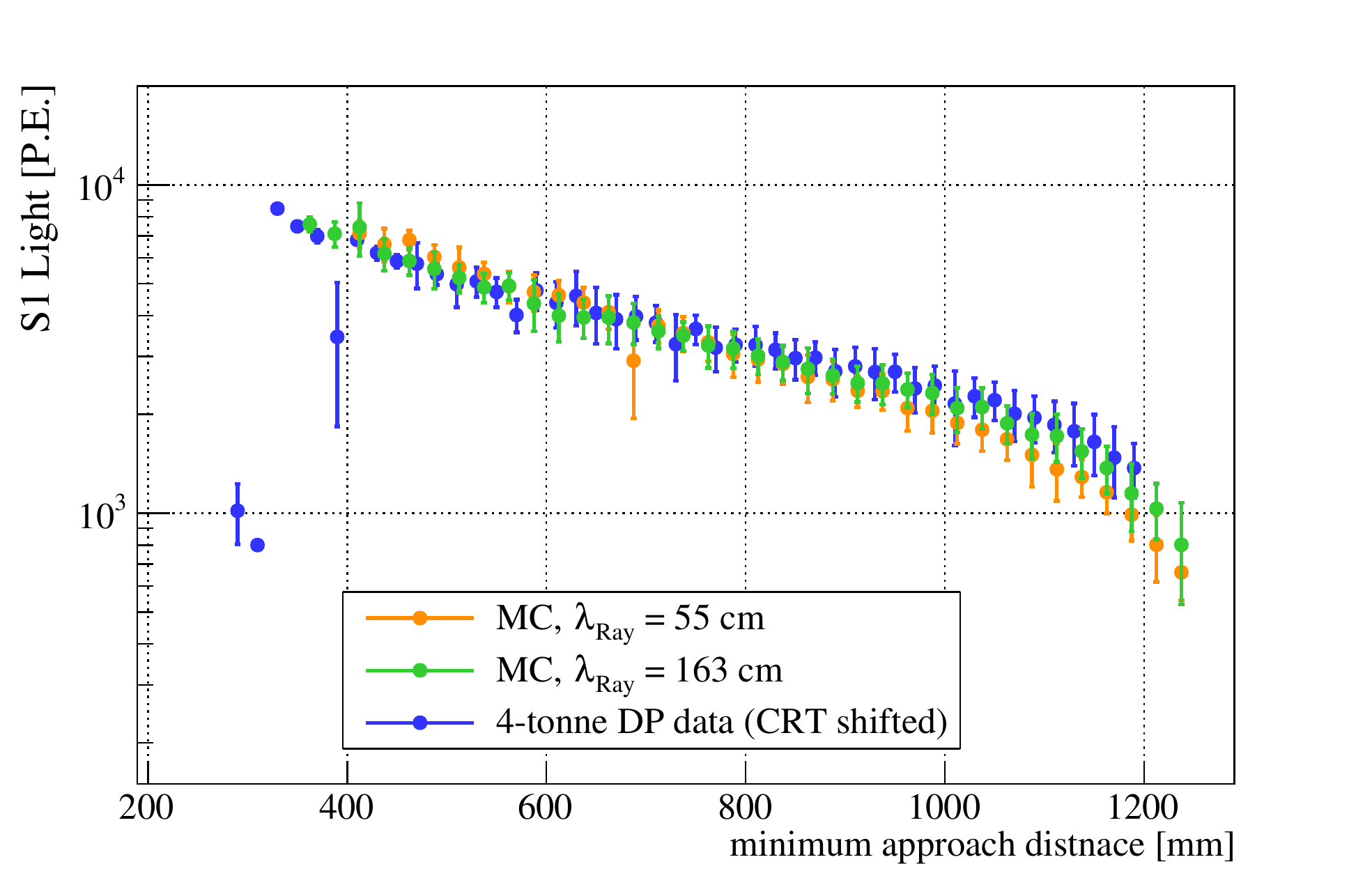}
 \caption{Study of the Rayleigh scattering length in the 4-tonne demonstrator. 
 %three different values of the Rayleigh scattering length parameter were simulated: 20 cm, 55cm and 163cm. The 4-tonne data are shown with the blue dots, the errors are statistical only.   
 Two different values of the Rayleigh scattering length parameter were simulated: 55cm and 163cm; the 4-tonne data are shown in blue.   
 \label{fig:311_results_Ray}}
\end{figure}

%The scintillation light in the 4-tonne demonstrator was simulated with a Geant4-based Monte Carlo (MC). The event generation used the muons kinematic measured by the CRTs. The NEST model \cite{NEST} was used to predict the amount of photons produced and the timing for each interaction. %The light propagation was done through light maps that assign to each photon the corresponding probability to be collected by the PMT units. The LAr and GAr phases are treated separately. 
The knowledge of light propagation in the LAr is not completely understood, hence the combination of data analysis with the MC simulation is crucial to investigate optical parameters. The simulation of the scintillation light in the 4-tonne demonstrator is based on the NEST model \cite{NEST} that predicts the amount of light produced and the timing for each interaction. The interactions are generated reflecting the muons kinematic measured by the CRTs. The light is propagated using 3D light collection maps that estimate and assign the number of photons detected by each PMT. The Rayleigh scattering and the absorption lengths are introduced as parameters of the collection probability assigned to each photon and stored in the 3D light maps. The propagation in the LAr and GAr phases is obtained from two separate maps\cite{AnnePhD}.\\
The Rayleigh scattering length was studied considering the correlation between the light collected by each PMT and the minimum approach distance between a muon track and a PMT photo-cathode. Different values of the Rayleigh scattering length are given in literature: 55cm and 163cm \cite{55cm}, \cite{163cm}, and were used in the simulations.
%In addition, a third very short value of 20 cm was simulated with a clear impact in the light propagation \cite{AnnePhD}. 
As shown in figure \ref{fig:311_results_Ray}, 
%our measurements are incompatible with $\lambda_{Ray}$=20 cm, as expected, but 
given the maximum PMT-track distances ($\sim$1m) 
%they 
our measurements are compatible with both $\lambda_{Ray}$=55 cm and $\lambda_{Ray}$=163 cm. %More studies are ongoing to quantify better this result.

\subsection{The scintillation light}
\label{sec:larlight}
%One of the two processes that can take place when a particle traverse the LAr volume is the formation of $\mathrm{Ar}^{*}_{2}$ excimers whose de-excitation produces scintillation light. 
The scintillation light is produced by the de-excitation of the $\mathrm{Ar}^{*}_{2}$ excimers ($\mathrm{Ar}^{*}_{2} \rightarrow \mathrm{Ar} + \mathrm{Ar} + \gamma$); their formation can be caused either by the direct excitation of Ar ($\mathrm{Ar}^{*} + \mathrm{Ar} \rightarrow \mathrm{Ar}^{*}_{2}$) or by the recombination of the $\mathrm{Ar}_{2}^{+}+e^{-}$ pairs produced by ionization ($\mathrm{Ar}^{+} + \mathrm{Ar} \rightarrow \mathrm{Ar}_{2}^{+}+e^{-} \rightarrow \mathrm{Ar}^{*}_{2}$). The characterization of the scintillation signal was possible fitting the time profile of the selected muons. The fit function is the convolution of a Gaussian, to model the detector response, with three exponentials:
\begin{equation}
\label{eq:scifit}
F(t) = G(t-t_{0},\sigma) \otimes \Sigma_{i=f,i,s} (A_{i}/\tau_{i})\times\exp(-(t-t_{0})/\tau_i)
\end{equation}
Two exponential functions are used to describe the \emph{fast} de-excitation from the singlet state ($\tau_{fast}\sim$6 ns), and the \emph{slow} de-excitation from the triplet state ($\tau_{slow}\sim$1.6 $\mu$s) \cite{Hitachi}. Given the measured scintillation time profile, an additional \emph{intermediate} component was needed to fit the transition region between the \emph{fast} and the \emph{slow} components. The origin of the \emph{intermediate} component has not been identified yet but various hypotheses have been proposed, one of those suggests that the TPB can be the cause of such third delayed re-emission \cite{Segreto}. %The value found for the \emph{intermediate} decay time in our measurement is $\sim$50 ns, on average. 
With our model, we didn't test directly any specific hypothesis on the origin of the \emph{intermediate} component, nevertheless, including the results of all the data collected, the value found for the \emph{intermediate} decay time is (50.3 $\pm$ 6.4) ns. The relative probability amplitude of each component is obtained from the normalization constants ($\mathrm{A_i}$), where i index refers to \emph{fast}, \emph{intermediate} and \emph{slow} components. 

The scintillation time profile is very sensitive to the detector conditions; in particular, the slow component is affected by the possible presence of impurities that quenches the scintillation light (e.g. $\mathrm{O}_2$ and $\mathrm{N}_2$) \cite{Acciarri}. Monitoring of the LAr purity is possible through the measurement of the $\tau_{slow}$ value. %In the 4-tonne demonstrator, this value remained stable during the whole data taking. The $\tau_{slow}$ was measured to be slightly higher than $\sim$1.4$\mu s$, corresponding to a very low amount of impurities ($<$ ppm level \cite{Acciarri}), in accordance with the charge measurement \cite{4TON}.
In the 4-tonne demonstrator, this value remained stable for more than five months of data taking. Including the results from all the runs collected in absence of drift field, $\tau_{slow}$ = (1.43 $\pm$ 0.04) $\mu$s is found, in agreement with \cite{Lippincott}. This value is consistent with the amount of impurities measured at the end of the cooling down phase by RGTA (in particular, [$\mathrm{O}_2] \sim$ 50 ppb), and with the electron lifetime measurement \cite{4TON}.

\begin{figure}[h!ptb]
  \centering 
  \includegraphics[width=0.485\linewidth,  height=5.3cm ]{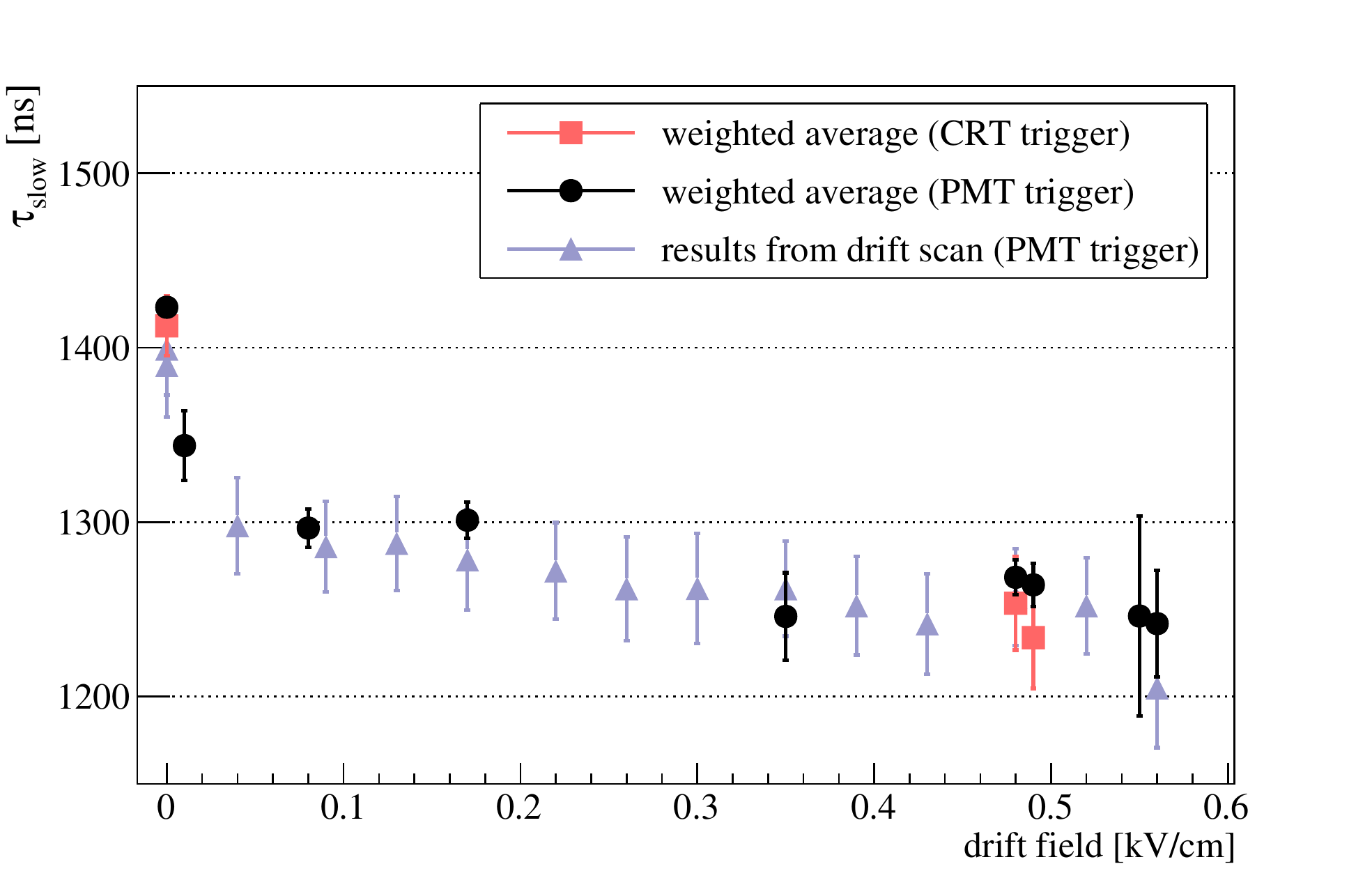} 
  \includegraphics[width=0.485\linewidth,  height=5.2cm ]{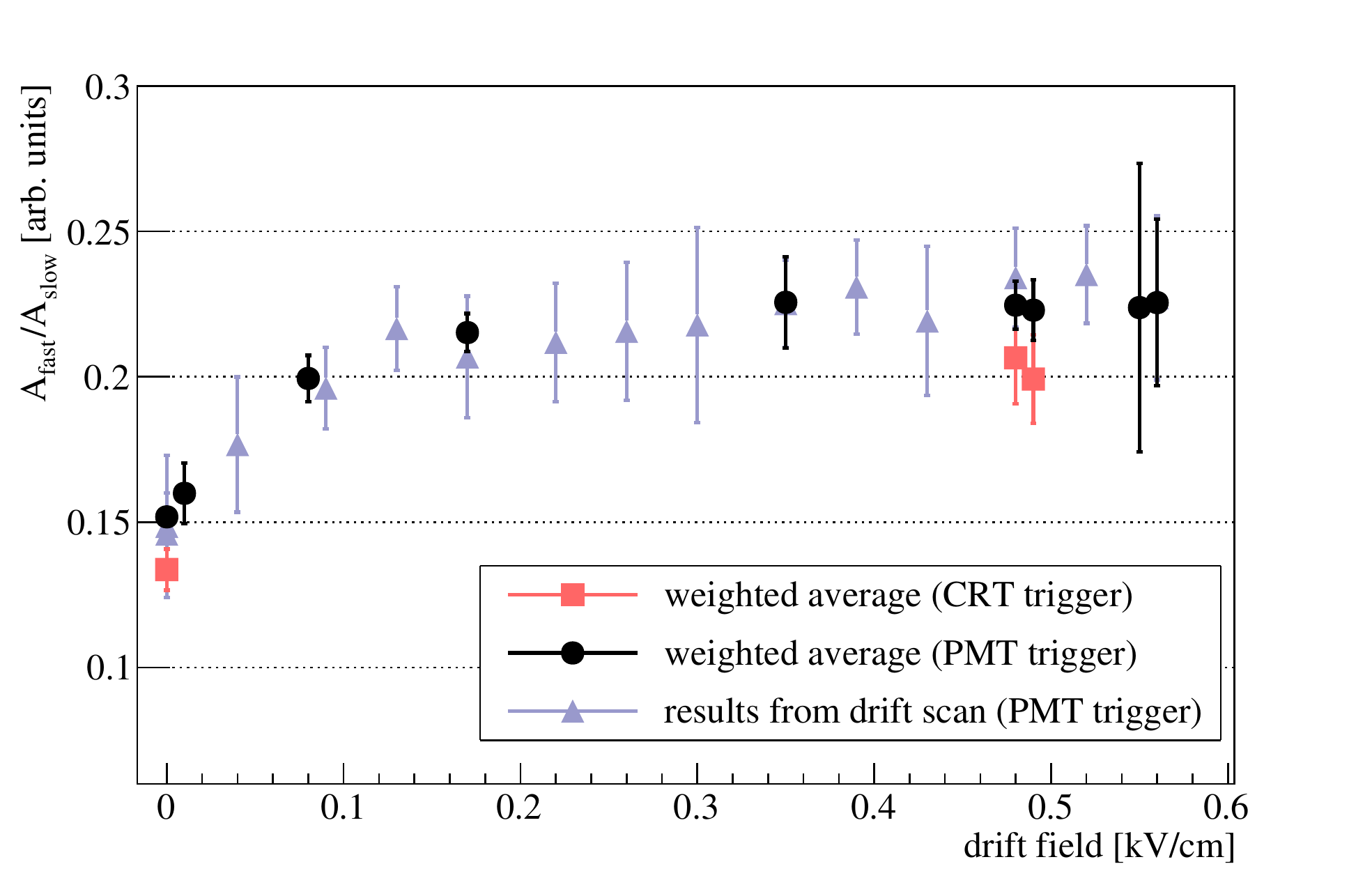} 
  \caption{Effect of the drift field on the scintillation time profile fit parameters. %obtained using Eq. \ref{eq:scifit}. 
  The $\tau_{slow}$ dependence with the drift field is shown on the \textbf{left} and the trend of the $\mathrm{A_f/A_s}$%fast over slow probability amplitudes .. trend
  ratio is shown on the \textbf{right}. \label{fig:311_results_a}}
\end{figure}

%The strength of the drift field tends to suppress the recombination process. Since also the amount of scintillation light and its time profile can be affected, a fit of Eq. \ref{eq:scifit} as a function of the drift field was performed. Data collected in a dedicated drift field scan in controlled purity condition and other data available at different values of the drift field have been used. From the data taken up to 0.5 kV/cm, the fraction of light produced by recombination has been measured showing that at least 40$\%$ of the scintillation light produced without drift field comes from the recombination process. The main results coming from the fit of Eq. \ref{eq:scifit} are summarized in figure \ref{fig:311_results_a}. 
The presence of the drift field tends to suppress the recombination process; consequently, the amount of scintillation light is expected to be reduced \cite{Birks}. At the nominal drift value (0.5 kV/cm), the fraction of light measured shows a decreasing of $\sim$ 40$\%$ compared with the light produced in the absence of such field. Additionally, from the comparison of the average waveforms collected increasing the drift field, an effect on the fit parameters is measured. The main results are summarized in figure \ref{fig:311_results_a}; the data analyzed includes results from a dedicated drift field scan collected under controlled purity conditions (in the range 0-0.56 kV/cm) and all other data available in the two trigger conditions.
The $\tau_{slow}$ values tend to decrease when the drift field increases. Given the theoretical description of the LAr excitation, no dependence of the $\tau_{slow}$ with the drift field is expected; nevertheless, the trend is robust with all the data analyzed and it does not show any evident connection with the trigger conditions. We have not find yet a valid explanation for this behavior, dedicated measurement should be carried out %in ProtoDUNE-DP 
to confirm our measurement. % since there are no equivalent studies in the literature to compare with. 
On the other hand, the two probability amplitudes of the de-excitation from the singlet, $\mathrm{A_f}$, and the triplet levels, $\mathrm{A_s}$, are expected to change for the effect of the drift field. The mean value of the ratio $\mathrm{A_f/A_s}$ tends to increase as a function of the drift field, as it is shown in figure \ref{fig:311_results_a} - right. Similar behavior has been observed considering the ratio $\mathrm{(A_f+A_i)/A_s}$. Our results are in direct contradiction with what has been reported in literature \cite{Kubota78} where LAr purity conditions may have been different. 

\begin{figure}[h!ptb]
  \centering 
  \includegraphics[width=0.485\linewidth,  height=5.2cm ]{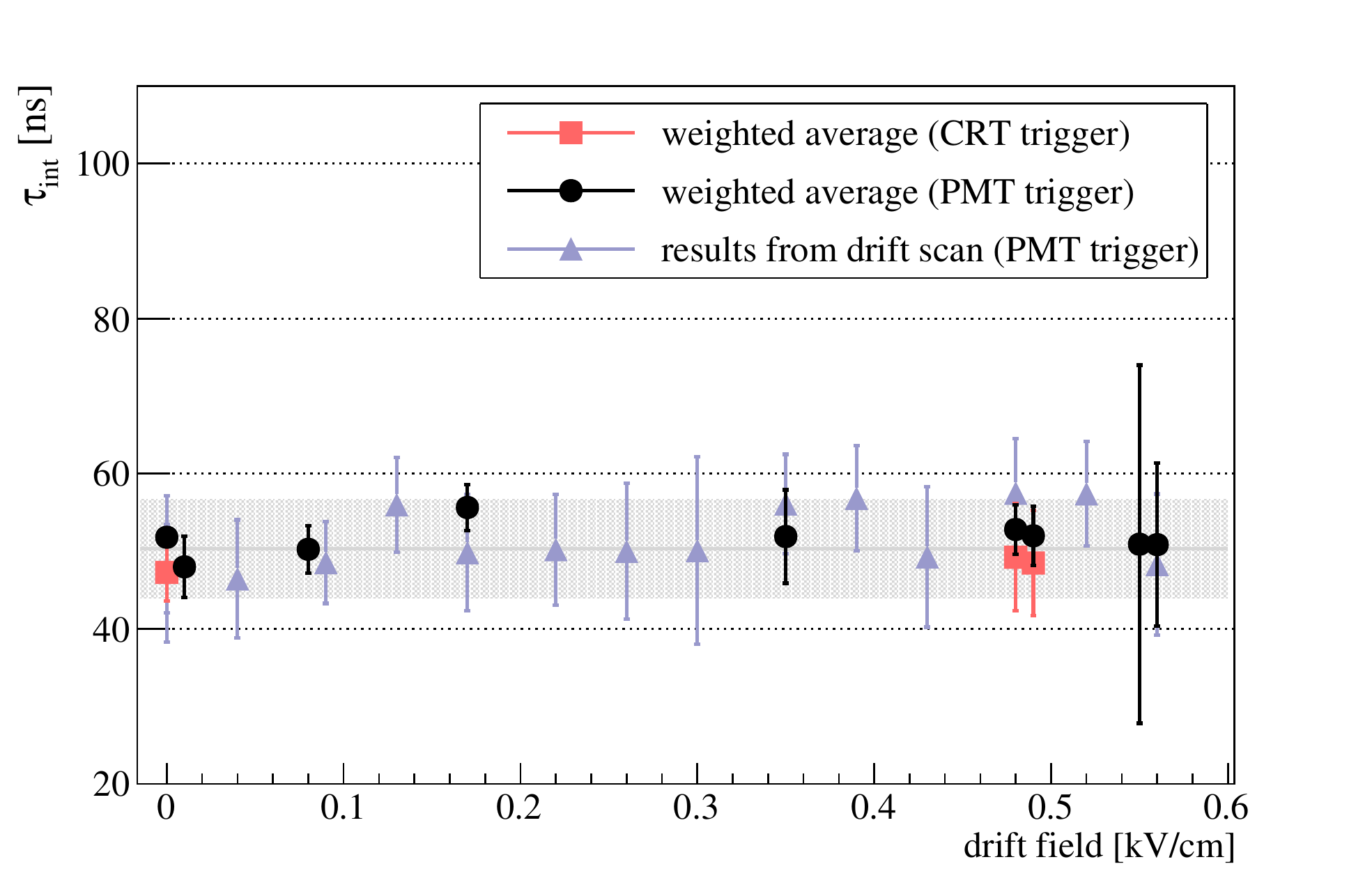} 
  \includegraphics[width=0.485\linewidth,  height=5.15cm ]{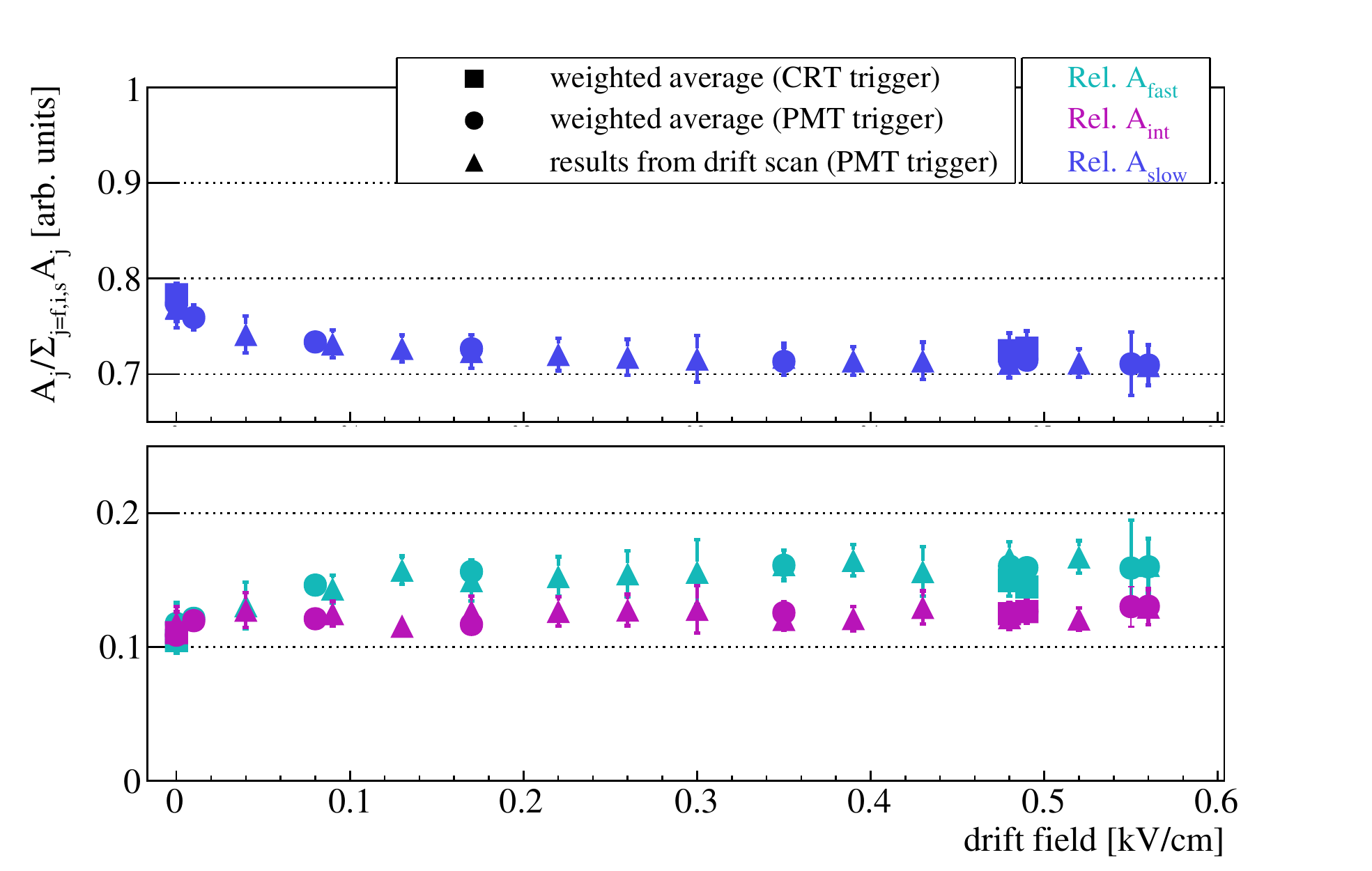} 
  \caption{Measurement of the decay time of the intermediate component; the average value measured is indicated by the gray band and corresponds to $\tau_{int}$ = (50.3 $\pm$ 6.4) ns.\label{fig:311_results_b}}
\end{figure}

The behavior of the \emph{intermediate} component as a function of the drift field is shown in figure \ref{fig:311_results_b}. Neither the $\tau_{int}$ nor the \emph{intermediate} probability amplitude show any dependence with the drift field. Other experiments \cite{Acciarri1}, report values of the ratio $\mathrm{(A_f + A_i)/A_s}$ measured in absence of drift field%. Our measurement is in agreement with this value
, however any dependence of the ratio $\mathrm{(A_f + A_i)/A_s}$ with the drift field is reported.% in these studies.

\subsection{The electro-luminescence light}
\label{sec:garlight}

The electro-luminescence light (S2) gives complementary information to the 2D charge track trajectory. For reconstructed muon-like events, it is possible to retrieve the S2 signal parameters as the time, the duration, the amplitude, and its delay with the S1 scintillation time peak, using a dedicated algorithm developed to reconstruct the S2 signal. All this information is related with the track topology and with the Ar properties affecting the drifted electrons. For instance,
the measurement of the electron drift velocity at $\sim$0.5 kV/cm was possible considering the correlation between the time difference $\Delta T_{\mathrm{S2-S1}}$ and the drift distance measured by the CRTs. Combining the results obtained from the five PMTs, a preliminary value of $\mathrm{v_{drift}} = $ (1.57 $\pm$ 0.06) mm/$\mu$s was measured; the error reported is purely statistical. %A better event selection is ongoing to improve this study. 
A better event selection is ongoing to improve this study, results from all the runs available will be included. 

\section{Conclusions and Prospects}
\label{sec:concl}
%The operation of the 4-tonne demonstrator had a crucial role in the scaling of the DP technology toward ProtoDUNE-DP. The light detection system fulfilled the basic requirement to be used as a self-trigger for the cosmic muons. The characterization and the analysis of the S1 and S2 light signals demonstrated the potential to use the scintillation light to obtain additional and complementary event information to the charge signal.

%Some of our new results are in contradiction with other measurements in the literature, emphasizing the importance to investigate light properties in LAr to take the maximum advantage of the information enclosed in the light signal in LAr-TPCs. Similar analyses will be carried out in ProtoDUNE-DP to improve and clarify these results.

The operation of the 4-tonne demonstrator had a crucial role in the scaling of the DP technology toward ProtoDUNE-DP. The light detection system fulfilled the basic requirement to be used as a self-trigger for the cosmic muons. The characterization and the analysis of the S1 and S2 light signals demonstrated the potential to use the scintillation light to obtain additional and complementary event information to the charge signal. The monitoring of the tau slow shows the stability of the LAr purity for more than five months, $\tau_{slow}$ = (1.43 $\pm$ 0.04) $\mu$s. An intermediate component is needed to fit the scintillation time profile and the corresponding decay time has been measured, $\tau_{int}$ = (50.3 $\pm$ 6.4) ns. The complementarity of the S2 time with the track topology has been used for a preliminary calculation of the electron drift velocity at the nominal drift field. 

Some of our new results, as the decreasing of the slow decay time and the increase of the ratio $\mathrm{A_f/A_s}$ for stronger drift fields, are in contradiction with other measurements in the literature; this emphasizes the importance to investigate light properties in LAr to take the maximum advantage of the information enclosed in the light signal in LAr-TPCs. Similar analyses will be carried out in ProtoDUNE-DP to improve and clarify these results.

%We suggest not to abbreviate: ``section'', ``appendix'', ``figure''
%and ``table'', but ``eq.'' and ``ref.'' are welcome. Also, please do
%not use \texttt{\textbackslash emph} or \texttt{\textbackslash it} for
%latin abbreviaitons: i.e., et al., e.g., vs., etc.

%\subsection{And subsequent}
%\paragraph{Up to paragraphs.} 

%\acknowledgments
%This is the most common positions for acknowledgments. A macro is
%available to maintain the same layout and spelling of the heading.

%\paragraph{Note added.} This is also a good position for notes added
%after the paper has been written.


%\clearpage

\end{document}